\documentstyle{mn}
\input{psfig}
\begin{document}
\def\ltsima{$\; \buildrel < \over \sim \;$}
\def\simlt{\lower.5ex\hbox{\ltsima}}
\def\gtsima{$\; \buildrel > \over \sim \;$}
\def\simgt{\lower.5ex\hbox{\gtsima}}
\def\approxgt{\mathrel{\hbox{\rlap{\lower.55ex \hbox {$\sim$}}
        \kern-.3em \raise.4ex \hbox{$>$}}}}
\def\approxlt{\mathrel{\hbox{\rlap{\lower.55ex \hbox {$\sim$}}
        \kern-.3em \raise.4ex \hbox{$<$}}}}

\title[The obscured AGN in NGC~5643]
{Unveiling the nature of the highly obscured AGN in NGC~5643 with XMM-Newton}

\author[M. Guainazzi et al.]
{M.~Guainazzi$^1$, P.~Rodriguez-Pascual$^1$, A.C.~Fabian$^2$, K.~Iwasawa$^2$, and G.~Matt$^3$ \\ ~ \\
$^1$European Space Astronomy Center, RSSD of ESA, VILSPA, Apartado 50727,
E-28080 Madrid, Spain
\\
$^2$Institute of Astronomy, Madingley Road, Cambridge, CB3 0HA
\\
$^3$Dipartimento di Fisica ``E.Amaldi", Universit\`a ``Roma Tre", Via della Vasca Navale 84, I-00146 Roma, Italy 
}

\maketitle
\begin{abstract}
We present
results from an XMM-Newton observation of the
nearby Seyfert~2 galaxy NGC~5643. The nucleus exhibits
a very flat X-ray continuum above 2~keV, together with a prominent
K$_{\alpha}$ fluorescent iron line. This indicates
heavy obscuration.
We measure an absorbing column density $N_H$ in the range
6--10$\times 10^{23}$~cm$^{-2}$, either directly covering
the nuclear emission, or covering its Compton-reflection.
In the
latter case, we might be observing a rather unusual geometry
for the absorber,
whereby reflection from the inner
far side of a torus is in turn obscured by its
near side
outer atmosphere.
The nuclear emission might be then
either covered by a Compton-thick
absorber, or undergoing a transient
state of low activity.
A second source (christened "X-1" in
this paper) at the outskirts of
NGC~5643 optical surface outshines the
nucleus in X-rays.
If belonging to NGC~5643, it is the third brightest ($L_X \sim 4 \times
10^{40}$~erg~s$^{-1}$) known Ultra Luminous X-ray source.
Comparison with past large
aperture spectra of NGC~5643 unveils dramatic X-ray spectral changes above
1~keV. We interpret them as due to variability of the
active nucleus {\it and} of source X-1
intrinsic X-ray powers by a factor $\ge$10 and 5, respectively.

\end{abstract}


\section{Introduction}
The X-ray spectra of Seyfert~2 galaxies exhibit
significant photoelectric absorption
(Warwick et al. 1989; Awaki et al. 1991;
Turner et al. 1997; Risaliti et al. 1999).
This evidence has been interpreted as
supporting the predictions of the Seyfert unification scenarios,
whereby ``type~2'' objects are seen at high inclination angles with
respect to an azimuthally-symmetric gas and dust structure (the
``torus''), which prevents us from directly observing the
nucleus and the Broad Line Regions. ASCA
observations, however, unveiled Seyfert~2s with peculiar X-ray
spectral properties: an inverted spectrum (energy
index, $\alpha < 0$); large Equivalent Width (EW) K$_{\alpha}$ iron
fluorescent lines (EW from a few hundreds to thousands eV);
very little or no evidence for photoelectric
absorption (Elvis \& Lawrence 1988; Koyama et al. 1989;
Ueno et al. 1994). This
phenomenology was interpreted as due to the column density of the absorber
in this subclass of Seyfert~2s being so large, that it
suppresses almost to invisibility the X-ray continuum, and enhances
the contrast between the iron feature and the underlying
continuum. When the
absorber column density exceeds $N_H \simeq
\sigma^{-1}_t = 1.5 \times 10^{24}$~cm$^{-2}$, the absorbing matter is
optically thick to Compton scattering, and the nuclear photons are
downscattered to energies where the photoelectric absorbing cross
section becomes dominant. The nuclear continuum
is then substantially
suppressed below 10~keV, and the nucleus can be seen only
along reflected and/or scattered optical paths, which do not intercept
the obscuring matter. Indeed, later
BeppoSAX observations, which extended the sensitive bandpass beyond
10~keV, confirmed this hypothesis by
detecting nuclear emission piercing through $\approxgt
2 \times 10^{24}$~cm$^{-1}$ absorbers (Matt et al. 1997b; Vignati et
al. 1999; Guainazzi et al. 2000). All the above justify the
nomenclature of ``Compton-thick'' given to this sub-class of
Seyfert~2s.

The X-ray spectra of the three closest AGN (Centaurus~A, NGC~4945, and
the Circinus~Galaxy) are absorbed by column densities larger than
$10^{23}$~cm$^{-2}$, and two of them are Compton-thick. This simple
fact suggests that Compton-thick Seyfert~2 galaxies may constitute a
large fraction of their parent population. Indeed, BeppoSAX
measurements of large samples of optically-selected Seyfert~2 galaxies
suggested that they may constitute a fraction as large as 50 per cent of the
total population (Maiolino et al. 1998, Risaliti et al. 1999),
at least in the local universe. It is hard to tell whether
the fraction of heavily obscured objects remains that large at
cosmological distances. 
If so, this may have important consequences for the history of accretion,
and for
the total energy budget ratio between accretion and
stellar light in the universe \cite{fabian99}.

We have started a program to study a complete, unbiased,
optically-defined sample of Compton-thick Seyfert~2 galaxies
\cite{risaliti99} with
XMM-Newton \cite{jansen01}, with the main goal of characterizing their
X-ray spectral properties in the hard X-ray regime. Preliminary
results on the whole sample are discussed by Guainazzi et
al. (2004). In this paper, we present the
observation of NGC~5643. It allows us to describe in details
the analysis method followed in the study of the sample.
The interest of this specific
observations is two-fold: a) the large collecting area
of the XMM-Newton optics
allows us to measure a, albeit
extreme, Compton-thin absorber;
b) better imaging resolution with respect to
previous mission allows us to discover a
serendipitous source, which apparently belongs to the optical/UV
surface of the host galaxy, and outshines the nucleus. If this source is
indeed associated with the NGC~5643 host galaxy, it represent one of
the brightest Ultra-Luminous X-ray (ULX) source ever observed, with
a total X-ray luminosity larger then $4 \times 10^{40}$~erg~s$^{-1}$.

\subsection{Some properties of NGC~5643}

NGC~5643 is a nearby ($z=0.004$) SAB(rs)C galaxy, known to host a
low-luminosity Seyfert~2 nucleus \cite{phillips83}.
It exhibits an extended emission line
region elongated in a direction close
to the radio position angle \cite{morris85},
due to a "V"-shaped structure of highly excited
gas. This is probably the projection
of a 1.8~kpc, one-sided ionization cone
\cite{simpson97}. Perpendicular to the radio
axis, a dust lane covers  the nucleus
\cite{simpson97}. The fact that NGC~5643 belongs to
the class of ``extreme infrared" galaxies
\cite{antonucci85} attracted much attention in the
past to try to explain the origin of IR
emission in this Active Galactic Nucleus (AGN).
Although intense
episodes of star formation are occurring in
its nearly circular arms, evidence
with respect to the nucleus is still
controversial. Mid-IR diagnostics suggest that the
AGN dominates the IR energy budget (Genzel et al. 1998).
Comparison of optical spectra
with synthesis models are, however, consistent with
a "starburst/Seyfert~2 composite" spectrum
\cite{cidfernandes01}. The NGC~5643 nucleus is
a strong radio emitter as well, most likely
powered by the AGN \cite{kewley00}.

The log of the observations discussed is
\begin{table}
\caption{Log of the observations discussed in this paper}
\begin{tabular}{lcc} \hline \hline
Observatory-instrument/s & Date & Exposure time \\
& & (ks) \\
ASCA-SIS/GIS & 21-Feb-1996 & 36.7/41.5 \\
BeppoSAX-LECS/MECS & 01-Mar-1997 & 7.1/10.4 \\
ROSAT-HRI & 28-Aug-1997 & 10.0 \\
XMM-Newton-MOS/pn & 08-Feb-2002 & 9.4/7.1 \\ \hline \hline
\end{tabular}
\label{tab9}
\end{table}
this paper is reported in Table~\ref{tab9}.
In this paper: energies are quoted in the source reference
frame; errors are at the 90 per cent confidence level for
1 interesting parameter for the best-fit model parameters,
and at the 1-$\sigma$ level for count rates;
chemical mixture follows the Anders \& Grevesse
(1989) measurement; $H_0 = 70$~km~Mpc$^{-1}$~s$^{-1}$
\cite{bennett03}.
At the NGC~5643 distance (16.9~Mpc), 1$\arcsec$ corresponds
to 82~pc.

\section{XMM-Newton observation}

XMM-Newton \cite{jansen01} observed the sky region
encompassing NGC~5643 on February 8, 2003. 
The X-ray imaging EPIC cameras
(pn, Str\"uder et al. 2001; MOS, Turner et al. 2001)
were operated in Full Frame Mode, with the {\sc
Medium} optical photons blocking filter. The
Optical Monitor (OM; Mason et al. 2001) was operated in standard
Image Mode with the UVW1 filter, sensitive in
the 2500-4000$\AA$ bandpass. Data reduction was performed
with version 5.4.1 of the {\sc SAS} software package
\cite{jansen01}.
We employed the most updated calibration files available at
the time the reduction was performed (June 2003).
None of the X-ray sources discussed
in this paper was detected by the high-resolution
spectroscopy camera, RGS \cite{derherder01}.
Standard data screening criteria were
applied in the extraction of scientific products.
The particle background stayed at a quiescent level \cite{lumb02}
during the whole observation, making
unnecessary any background-rejection filtering.
Pattern 0 to 4 (12) were employed
in the extraction of pn (MOS) scientific products.
Spectra and light curves were extracted from circular
regions of 25$\arcsec$ radius around the source
centroids. Response matrices, appropriate for
each of the spectra discussed in this paper,
were generated with the {\sc SAS} tasks
{\tt arfgen} and {\tt rmfgen}.
Background spectra were extracted from
source-free regions of the same CCD as where the source is located.
Spectra were rebinned
to ensure that: a) the instrumental energy resolution
is oversampled by a factor not larger than 3;
b) at least 25 (50) counts are present in each
MOS (pn) spectral channel, to ensure the applicability
of the $\chi^2$ statistics in the evaluation of the fit
quality.
Images, spectra and response matrices of the two MOS
cameras were combined together, to increase
signal-to-noise. Spectral fits were performed
simultaneously on both EPIC cameras, using the
0.35--12 and 0.5--10~keV energy bands for the pn and the MOS,
respectively. As no significant variability was
detected during the XMM-Newton observation, in this
paper we
will discuss the time-averaged spectra only.

\begin{figure}
\begin{center}
\psfig{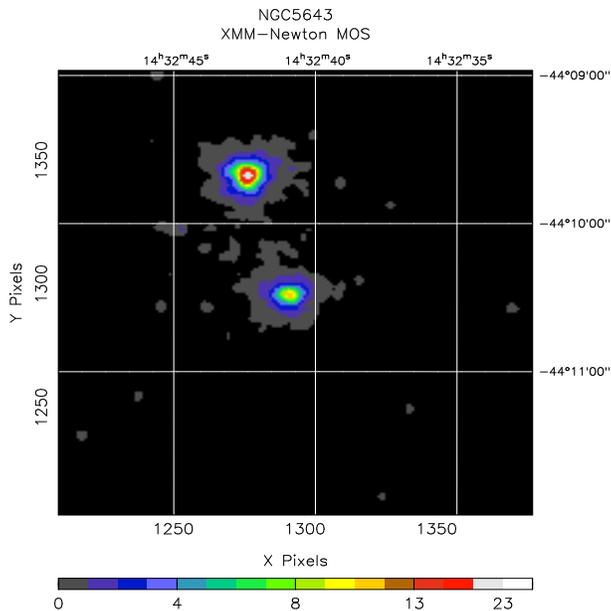}
\end{center}
\vspace{-0.5cm}
\caption{
0.5--10~keV single- and double-events MOS image,
restricted to the 3$\arcmin$ around the NGC~5643
optical nucleus. The image is smoothed
with a Gaussian function ($\sigma = 1.5\arcsec$).
The scale in is units of unsmoothed pixel counts.
}
\label{fig1}
\end{figure}

In Fig.~\ref{fig1} the
0.5--10~keV sky coordinates
MOS image of the innermost 3$\arcmin$ around
the NGC~5643 nucleus is shown. Four sources are detected
by the EPIC cameras
in this field at a signal-to-noise ratio larger than
3. The source detected at
($\alpha_{2000}=14^h32^m40.9^s$, $\delta_{2000}=-44^{\circ}10{\arcmin}26{\arcsec})$
coincides with NGC~5643 optical nucleus within the typical
XMM-Newton attitude reconstruction accuracy. Its total
counts are $1380 \pm 40$ and
$640 \pm 30$ in the pn and MOS camera, respectively.
The source
located $\simeq$0.8$\arcmin$NE the nucleus
is 50 per cent brighter then
the nucleus. We will refer to it as ``NGC~5643 X-1'' hereafter.
Its total counts are $2030 \pm 60$ and $1320 \pm 40$ in the
pn and MOS cameras, respectively.

\subsection{Spectral analysis of the nucleus}

\begin{figure}
\begin{center}
\psfig{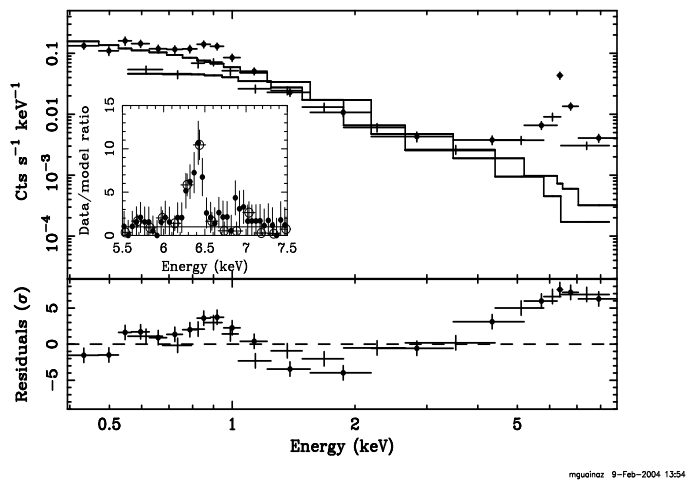}
\end{center}
\vspace{-0.5cm}
\caption{
NGC~5643 nucleus EPIC spectra ({\it upper panel};
{\it dots}: pn; {\it crosses}: MOS) and
residuals in units of standard deviations ({\it lower
panel}) against a power-law model. In the {\it inset}:
data/model ratio in the 5.5-7.5~keV energy band against
a power-law model above 2~keV (pn: {\it filled circles};
MOS {\it empty squares}). In the latter panel, data
are rebinned according to a constant factor $\delta
E = 50$~eV for plotting purposes only
}
\label{fig3}
\end{figure}

In Fig.~\ref{fig3} we show the
result of a power-law fit\footnote{All the models
employed in this paper include photoelectric
absorption by a column density due to the contribution
of neutral matter in our Galaxy along the
line-of-sight to NGC~5643: $N_{H,Gal} = 8.3 \times
10^{20}$~cm$^{-2}$ (Dickey \& Lockman 1990)}
on the NGC~5643 nucleus spectrum. The fit
is clearly unacceptable ($\chi^2/\nu = 417.0/29$).
As often observed in highly absorbed Seyfert~2
galaxies \cite{matt00} the residuals exhibit a
``turning point'' around 2~keV, suggesting that
the continuum needs to be modeled with (at least)
two components. Additionally, a broad excess emission
feature is localized around $E \simeq 0.85$~keV
(observer's frame), together with a narrow emission
feature around 6~keV. In order to achieve a full
physical characterization of the overall spectrum,
we have first separately analyzed the regimes
above and below the spectral ``turning point''.

\subsubsection{The soft X-ray spectrum ($E < 2$~keV)}

The soft energy spectrum is remarkably complex.
A fit with a single featureless continuum
in the energy band redwards of 1~keV is clearly inadequate
({\it e.g.}: $\chi^2_{\nu} \simeq 3.5$ if a power-law is
employed), and leaves large positive residuals in
the energy range between 0.8 and 1~keV. In 
this region a ``forest'' of iron-L transitions
exists. At
least two thermal components\footnote{hereafter thermal emission from
an optically thin, collisionally excited
plasma is modeled via the code {\tt mekal}
in {\sc Xspec} \cite{mewe85}. Unless
otherwise specified, the plasma abundance has
been kept fixed to $Z = 0.5 Z_{\odot}$} are required
in order to adequately account for this feature,
with temperatures
$kT_s \simeq 0.17$~keV
and $kT_h \simeq 0.69$~keV
($\chi^2/\nu = 9.7/8$).
Alternative
``two continuum components''
descriptions of the soft X-ray spectrum are possible,
and yield statistically equivalent fit,
such as the combination of
thermal emission and power-law ($kT \simeq
1.1$~keV; $\Gamma \simeq 2.2$), or of
thermal and multi-temperature accretion disk
blackbody emission (model {\tt diskbb} in
{\sc Xspec};
$kT^{thermal} \simeq 90$~eV; $kT^{disk} \simeq 1.0$~keV).

\begin{figure}
\begin{center}
\psfig{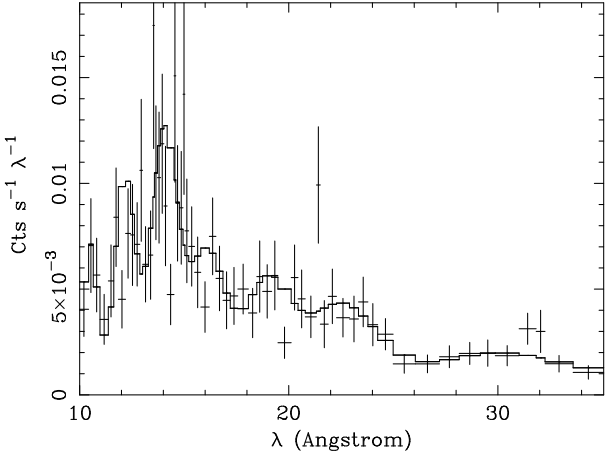}
\end{center}
\vspace{-0.5cm}
\caption{
10--35$\AA$ pn spectrum ({\it crosses}) superposed to the
emission-line only model best-fit ({\it solid line} - details
in text)
}
\label{fig8}
\end{figure}

Is it alternatively
possible that the soft X-ray spectrum is dominated
by scattering? The soft
X-ray spectrum can be in principle reasonably fit
with a combination of a, say, power-law continuum
plus three emission lines ($\chi^2_{\nu} \simeq 1.23$).
However the spectral index in this scenario
($\Gamma \simeq 4.0$) is too steep to
represent the mirror of the AGN intrinsic continuum, even if
self absorption effect in the scattering plasma are
neglected. However, high-resolution grating measurements
of bright absorbed Seyfert galaxies are showing that
the soft X-ray emission is often dominated
by heavily blended emission lines
(Kinkhabwala et al. 2002; Brinkmann et al. 2002;
Sako et al 2000). The blend can mimic
continuum emission - alongside a few emission
``peaks" corresponding to the brightest well defined lines -
in low-resolution spectra
(Guainazzi et al. 1999; Cappi et al. 1999).
Testing this hypothesis would require high-resolution
good quality spectroscopic data, which are not available
for the NGC~5643 nucleus. Nonetheless, we
fit the EPIC spectra with a combination of emission
lines only, aiming at inferring the main qualitative
properties of the line emitting plasma in this scenario.
In order to
avoid line blurring effects induced by spectral
rebinning, we fit the unbinned spectrum, using the
C-statistics \cite{cash76}. Only the pn spectrum
was used, due to
larger effective area. No
background subtraction was performed, as its
contribution is $\le 10$~per cent in the soft X-ray band. Seven lines
are required to fit the spectrum
(see Fig.~\ref{fig8}, and
Table~\ref{tab10}). Typical statistical uncertainties
\begin{table}
\caption{Centroid energies for the measured emission lines,
if the soft X-ray pn spectrum of the NGC~5643 nucleus
is fit with a model constituted by emission lines only. The
{\it right column} shown the possible identifications.
They take into account the statistical uncertainties
on the centroid energies only. In brackets,
additional identifications are listed when systematics
uncertainties are taken into account as well.
RRC~=~Radiative Recombination Continuum}
\begin{tabular}{cl} \hline \hline
Energy (keV) & Identifications \\ \hline
0.428 & C{\sc vi} $\beta$ (N{\sc vi})\\
0.566 & O{\sc vii} \\
0.668 & O{\sc vii} $\gamma$ \\
0.787 & Fe{\sc xvii} \\
0.899 & Ne{\sc ix}, Fe{\sc xix} \\
1.028 & Ne{\sc ix} $\beta$, Fe{\sc xxii} \\
1.176 & Ne{\sc ix} RRC, Fe{\sc xxiv} \\ \hline \hline
\end{tabular}
\label{tab10}
\end{table}
on their centroid energy
are $\simeq 10$~eV (except for the
highest energy line in Table~\ref{tab10}, for which they
are $\simeq 20$~eV). Systematic uncertainties on the
gain are of the same order.

Better data quality
is necessary to achieve a fully unambiguous
deconvolution of the soft X-ray spectrum.
For sake of simplicity, we will assume in the following
a modeling of the soft X-ray spectrum in terms of
a two-component optically-thin thermal plasma.

\subsubsection{The hard X-ray spectrum ($E > 2$~keV)}

The nuclear spectrum above 2~keV is no less complex.
A formally adequate
fit is achieved through an unabsorbed power-law and
an emission line (parameterized through a simple
Gaussian profile). The latter is required
at the 99.87 per cent confidence level
according to the F-test. However, the best-fit parameters
suggest that this is no more than a purely mathematical
description. The power-law spectral index
is inverted
(photon index, $\Gamma \simeq -0.8$), what is totally
unusual for AGN (Nandra et al. 1997;
et al. 1997; Reeves \& Turner 2000; Perola et al. 2002).
The emission line
centroid energy ($E_c \simeq 6.42$~keV) is
consistent with K$_{\alpha}$ fluorescent
emission from neutral or mildly ionized iron.
The formal
line Equivalent Width is huge: $EW \simeq 2.8$~keV
(cf. the inset in Fig.~\ref{fig3}).

Flat hard X-ray spectra
with huge EW K$_{\alpha}$ fluorescent iron lines
in Seyfert~2 galaxies can be produced if
the bulk of hard X-ray emission
is due to Compton-reflection of an
otherwise invisible ``standard'' AGN nuclear
continuum, either because
the AGN switched to an ``off'' state during
the XMM-Newton observation, or because its emission is
totally suppressed by a Compton-thick absorber
\cite{matt03}. We have therefore fit the 
EPIC nuclear spectra
in their full energy bandpasses with a combination of
two thermal components, a Gaussian emission
line, and a ``bare'' Compton-reflection component
(model {\tt pexrav} in {\sc Xspec}; Magdziarz \&
Zdziarski 1995). The fit is unacceptable
($\chi^2/\nu = 49.9/22$), mainly due
to large excess residuals between 2 and
5~keV. This excess could be in principle
due to optically thin
electron scattering of the otherwise invisible
nuclear continuum. Indeed, adding a power-law
component yields a statistically acceptable fit
($\chi^2/\nu = 18.6/21$). However,
the spectral index of this power-law is too flat
($\Gamma = -0.1 \pm^{0.8}_{0.6}$),
for its interpretation as the ``warm mirror'' of the
nuclear continuum to be plausible. The combination
of a Compton-thick absorber with nuclear electron
scattering can be therefore ruled out.

A possible alternative is that the Compton-reflection component
is in turn absorbed by matter with a covering fraction lower
than 1. In this case, the
2--5~keV spectrum is accounted for by $\simeq$5\% of the
Compton-reflection continuum (see Model\#1 in Table~\ref{tab2})
which ``leaks'' through its absorber. This leakage has
the appropriate flatness.
An absorber with a $\simeq$95~per cent covering fraction
and $N_H = (1.0 \pm 0.3) \times 10^{24}$~cm$^{-2}$ is
required in this scenario.
If the hard X-ray continuum is dominated by Compton-reflection,
K$_{\alpha}$ emission lines from O to Cr are expected to
imprint their signatures along with the Fe line. They
are not detected in the EPIC spectra of NGC~5643.
However, their EW upper limits (200-300)~eV are inconclusive
\cite{matt97b}.

The depth of the K photoelectric absorption edge is
in principle a powerful diagnostic for the metallicity
of the absorbing and/or reflecting matter. A
further 97~per cent confidence
level improvement in the
quality of the fit ($\Delta \chi^2/\Delta \nu = 5.0/1$)
is obtained in Model\#1 of Table~\ref{tab2}
if the iron metallicity is left free. If
this change is attributed
to one of the two components only (what is unlikely,
if absorber and reflector represent one and the same
system; see Sect.~4.1): $Z_{Fe} \le 0.7 Z_{\odot}$.

Alternatively, Compton-thin obscuration - still allowing
a fraction of the direct nuclear emission to pierce
through the X-ray absorber in the EPIC energy bandpass -
can as well significantly harden a typical AGN spectrum.
We have therefore substituted the
partially covered Compton-reflection
component with a power-law modified by photoelectric
absorption.
Again the fit is statistically acceptable
($\chi^2_{\nu} \simeq 0.95$). The best-fit models
results are reported
in Table~\ref{tab2}.
\begin{table}
\caption{Best-fit parameters and results for the
EPIC NGC~5643 nuclear spectrum.
Model \#1 is constituted by
the combination of two thermal components
(with temperatures $kT_s$, and $kT_h$), a
Compton-reflection continuum, partly
absorbed by matter with a covering
fraction $f_c$ and column density $N_H$,
and a Gaussian emission line profile
with centroid energy $E_c$, intensity
$I_{line}$, and Equivalent Width $EW$. Model \#2 is
constituted by the combination of three thermal
components (the hottest with temperature
$kT_{uh}$), an absorbed
power-law, and the Gaussian emission line profile.
In Model \#3 the ``ultra-hard'' thermal component
is substituted by a locally unabsorbed power-law.
$f_s$ is defined as the scattering fraction
between the transmitted and the scattered nuclear
component in Model \#3}
\begin{tabular}{lccc} \hline \hline
Parameter & Model \#1 & Model \#2 & Model \#3 \\ 
$\Gamma$ & $2.3\pm^{0.9}_{1.0}$ & $1.6 \pm^{1.0}_{1.9}$ & $1.4 \pm^{0.7}_{0.2}$ \\
$N_H$ ($10^{23}$~cm$^{-2}$) & $10 \pm 3$ & $6 \pm 4$ & $7 \pm 2$ \\
$f_c$ or $f_s$ (per cent) & $94.7 \pm_{5.6}^{3.4}$ & ... & $3.3 \pm^{7.3}_{2.8}$ \\
$E_c$ (keV) & $6.39 \pm^{0.04}_{0.05}$ & $6.43 \pm^{0.02}_{0.05}$ & $6.43 \pm^{0.02}_{0.05}$ \\
$I_{line}$$^a$ & $8 \pm 4$ & $1.4 \pm 0.4$ & $1.5 \pm^{0.4}_{0.3}$ \\
$EW$ (eV) & 760 & 490 & 500 \\
$kT_s$ (keV) & $0.17 \pm^{0.07}_{0.08}$ & $0.15 \pm^{0.03}_{0.07}$ & $0.15 \pm^{0.03}_{0.07}$ \\
$kT_h$ (keV) & $0.73\pm^{0.10}_{0.08}$ & $0.67 \pm 0.05$ & $0.67 \pm^{0.05}_{0.04}$ \\
$kT_{uh}$ (keV) & ... &  $> 5$ & ... \\
0.5--2~keV flux$^b$ & ... & ... & $2.16 \pm 0.15$ \\
2--10~keV flux$^b$ & ... & ... & $8.4 \pm^{1.3}_{1.2}$ \\
$\chi^2/\nu$ & 21.2/19 & 20.0/21 & 20.8/22 \\ \hline \hline
\end{tabular}
\noindent
$^a$in units of $10^{-5}$~cm$^{-2}$~s$^{-1}$

\noindent
$^b$corrected for Galactic absorption,
in units of $10^{-13}$~erg~cm$^{-2}$~s$^{-1}$
\label{tab2}
\end{table}
In this scenario the 2--5~keV excess can be accounted for
by optically-thin scattering of the nuclear continuum,
with a reasonable spectral shape ($\Gamma = 1.4 \pm^{0.7}_{0.2}$). 
The extrapolation
of the scattered nuclear continuum into the 0.4--1.2~keV
band is $\simeq$15~per cent in Model \#3.
No
significant evidence is found for a deviation of the absorber
metallicity 
from standard solar abundances in this scenario:
$Z_{Fe} = 0.8 \pm^{0.6}_{0.2} Z_{\odot}$.
In Model\#2 of Table~\ref{tab2} the electron scattering is substituted
by a ``ultra-hot'' ($kT \simeq 5$~keV) thermal component.

\begin{figure}
\begin{center}
\psfig{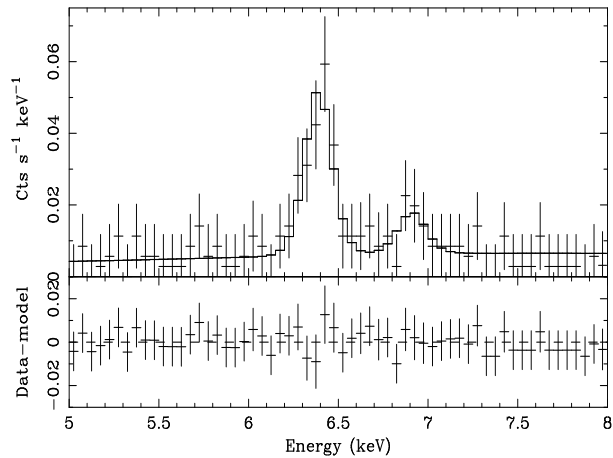}
\end{center}
\vspace{-0.5cm}
\caption{
{\it Upper panel}: pn spectrum ({\it crosses}) and best-fit
model ({\it solid line}) in the 4--8~keV energy
band, showing two emission lines with centroid energies
$\simeq 6.40$~keV and $\simeq 6.93$~keV. {\it Lower
panel}: residuals against the best-fit model.
A constant $\delta E = 50$~eV rebinning has been
applied for plotting purposes only
}
\label{fig9}
\end{figure}

In order to investigate the details
of the iron line structure, we have performed a Cash
statistic fit of
the unbinned pn spectrum in the 4--8~keV
energy range, again to prevent the line profile
from being blurred by the $\chi^2$ applicability
requirement. The contribution of the background
has been estimated by fitting the background
spectrum in the same energy band with a
simple power-law. The best-fit parameters
for the background are: $\Gamma = 3.08$ and
1~keV normalization,
$N = 5.8 \times 10^{-5}$~cm$^{-2}$~s$^{-1}$.
The source spectrum continuum was than fitted
with a double power-law, one of the
two components having its parameters fixed to
the background best-fit model. Two lines
are detected in the spectrum, at a
confidence level larger than 99~per cent for 1
interesting parameter (see Fig.~\ref{fig9};
Lampton et al. 1976).
They have centroid
energies $E_1 = 6.40 \pm 0.02$~keV and
$E_2 = 6.93 \pm 0.05$~keV, and intensities
$(1.4 \pm^{0.4}_{0.3}) \times 10^{-5}$~cm$^{-2}$~s$^{-1}$,
and $(4 \pm^3_2) \times 10^{-6}$~cm$^{-2}$~s$^{-1}$,
respectively. The fainter line is
consistent with K$_{\alpha}$ fluorescence of
H-like iron.

The observed fluxes - calculated according to
Model \#3
best-fit parameters
in Table~\ref{tab2} - are $1.7 \times 10^{-13}$ and
$8.3 \times 10^{-13}$~erg~cm$^{-2}$~s$^{-1}$
in the 0.5--2~keV and 2--10~keV energy ranges,
respectively. The corresponding luminosity in
the 0.5--2~keV energy range,
once corrected for
the Galactic absorption, is
$\simeq 9 \times 10^{39}$~erg~s$^{-1}$.
The intrinsic AGN luminosity, further corrected
for its local absorption, is
$(2.5 \pm^{7.2}_{2.0}) \times 10^{41}$~erg~s$^{-1}$ in the
0.5--10~keV energy range.
 
\subsection{Spectral analysis of NGC~5643 X-1}

\begin{figure}
\begin{center}
\psfig{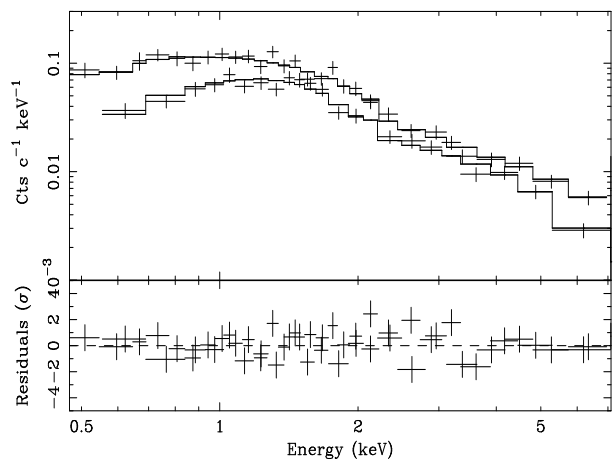}
\end{center}
\vspace{-0.5cm}
\caption{
Spectra ({\it upper panel}) and residuals in units of
standard deviation ({\it lower panel}), against
a photoelectrically
absorbed power-law for NGC~5643 X-1
}
\label{fig4}
\end{figure}

In Fig.~\ref{fig4} we show the EPIC spectra of NGC~5643 X-1
and its best-fit with a photoelectrically absorbed
power-law model. This fit is acceptable ($\chi^2/\nu = 46.4/48$).
An equally good fit can be obtained with a
thermal bremsstrahlung
continuum. A fit with a multitemperature blackbody
({\tt diskbb} in {\sc Xspec})
is instead poor.
A summary of the best-fit parameters and results is shown
in Table~\ref{tab3}. No iron K$_{\alpha}$ fluorescent
\begin{table}
\caption{Best-fit parameters and results for the
NGC~5643 X-1 EPIC spectra. Models legenda: {\tt po}~=~power-law;
{\tt diskbb}~=~multitemperature disk blackbody;
{\tt brems}~=~thermal bremsstrahlung,}
\begin{tabular}{lccc} \hline \hline
Parameter & {\tt po} & {\tt diskbb} & {\tt brems} \\
$N_H$($10^{21}$~cm$^{-2}$)& $1.7 \pm 0.4$ & $<$0.05 & $1.1 \pm^{0.2}_{0.3}$ \\
$\Gamma$ & $1.69 \pm^{0.09}_{0.10}$ & ... & ... \\
$kT$ (keV) & ... & $1.42 \pm^{0.08}_{0.09}$ & $8.5 \pm^{2.2}_{1.7}$ \\
0.5--2~keV flux$^a$ & $4.4 \pm 0.5$ & ... & ... \\
2--10~keV flux$^a$ & $7.5 \pm 0.6$ & ... & ... \\
$\chi^2/\nu$ & 46.3/48 & 87.8/48 & 45.9/48 \\
\hline \hline
\end{tabular}

\noindent
$^a$absorption-corrected,
in units of $10^{-13}$~erg~cm$^{-2}$~s$^{-1}$
\label{tab3}
\end{table}
emission line (or any other narrow band features) is detected
in the spectrum. The upper limit on the EW of a ``narrow''
({\it i.e.} $\sigma \equiv 0$), neutral ($E_c \equiv 6.4$~keV
in the observer's frame)
line is 270~eV.

The observed flux in the 0.5--10~keV energy band is
$(1.11 \pm 0.06) \times 10^{-12}$~erg~cm$^{-2}$~s$^{-1}$. It corresponds
to an intrinsic luminosity of $(4.4 \pm 0.2) \times
10^{40}$~erg~s$^{-1}$
if NGC~5643 X-1 is located at the galaxy redshift.

\section{The X-ray history of NGC~5643}

In this Section we compare the results of the XMM-Newton
observation of the NGC~5643 field with
earlier ASCA, BeppoSAX, and ROSAT/HRI observations.
Data were retrieved from the
NASA/HEASARC and ASI/ASDC archival
facilities, and reduced according to standard procedures.
In the HRI image both NGC~5643 XMM-Newton
sources are individually detected,
thanks to its good spatial resolution (see Table~\ref{tab7})
\begin{table}
\caption{Sources detected in the ROSAT/HRI observation
of the NGC~5643 field, within 3$\arcmin$ the optical
nucleus at a signal-to-noise ratio larger than 3}
\begin{tabular}{cccc} \hline \hline
$\alpha_{2000}$ & $\delta_{2000}$ & Count rate & Flux$^a$\\
& & ($10^{-3}$~s$^{-1}$) & \\ 
$14^h32^m40.7^s$ & $-44^{\circ}10{\arcmin}24{\arcsec}$ & $7.1 \pm
1.0$ & $2.3 \pm 0.3$ \\
$14^h32^m41.9^s$ & $-44^{\circ}09{\arcmin}36{\arcsec}$ & $3.2 \pm 0.7$ &
$1.5 \pm 0.3$ \\ \hline \hline
\end{tabular}

\noindent
$^a$0.5--2~keV flux
in units of $10^{-13}$~erg~cm$^{-2}$~s$^{-1}$,
absorption-corrected and extrapolated from
the measured count rate via the HEASARC on-line tool {\sc Pimms}, using the
best-fit models of the XMM-Newton observation
\label{tab7}
\end{table}
On the other hand, the broad Point Spread Function 
of the ASCA and BeppoSAX optics mixes irremediably
the contribution of the two sources in their
typical aperture. In both latter cases, we analyzed
a single spectrum extracted from a circular region of
3$\arcmin$ radius (except for the BeppoSAX LECS detector,
for which an aperture of 2$\arcmin$
was used due to the lack of appropriate calibrations).
Spectra were rebinned according to the same
criteria employed for the XMM-Newton/MOS spectra (cf.
Sect.~2.1). Background spectra were extracted
from nearby regions in the same field-of-view
where the source is located.
The
soft flux exhibits a dynamical range
of about a factor 2 between the
ASCA (faintest) and the XMM-Newton (brightest) state.
The variation amplitude in the hard X-rays is
somewhat smaller ($\simeq 30$~per cent).

Is it possible to
unambiguously ascribe the observed historical flux variability
either to the
NGC~5643 nucleus or to source X-1?  The fainter states
caught by BeppoSAX and - above all - ASCA
may represent a ``diluted'' version of a
more dramatic variability affecting only one
of the members of the
pair. Indeed, NGC~5643 X-1 underwent a factor of 3
variation in flux between the 1997 ROSAT/HRI
and the 2002 XMM-Newton observations. However, a quantitative
assessment of this issue across the whole X-ray
history of NGC~5643 is not straightforward.

\begin{figure*}
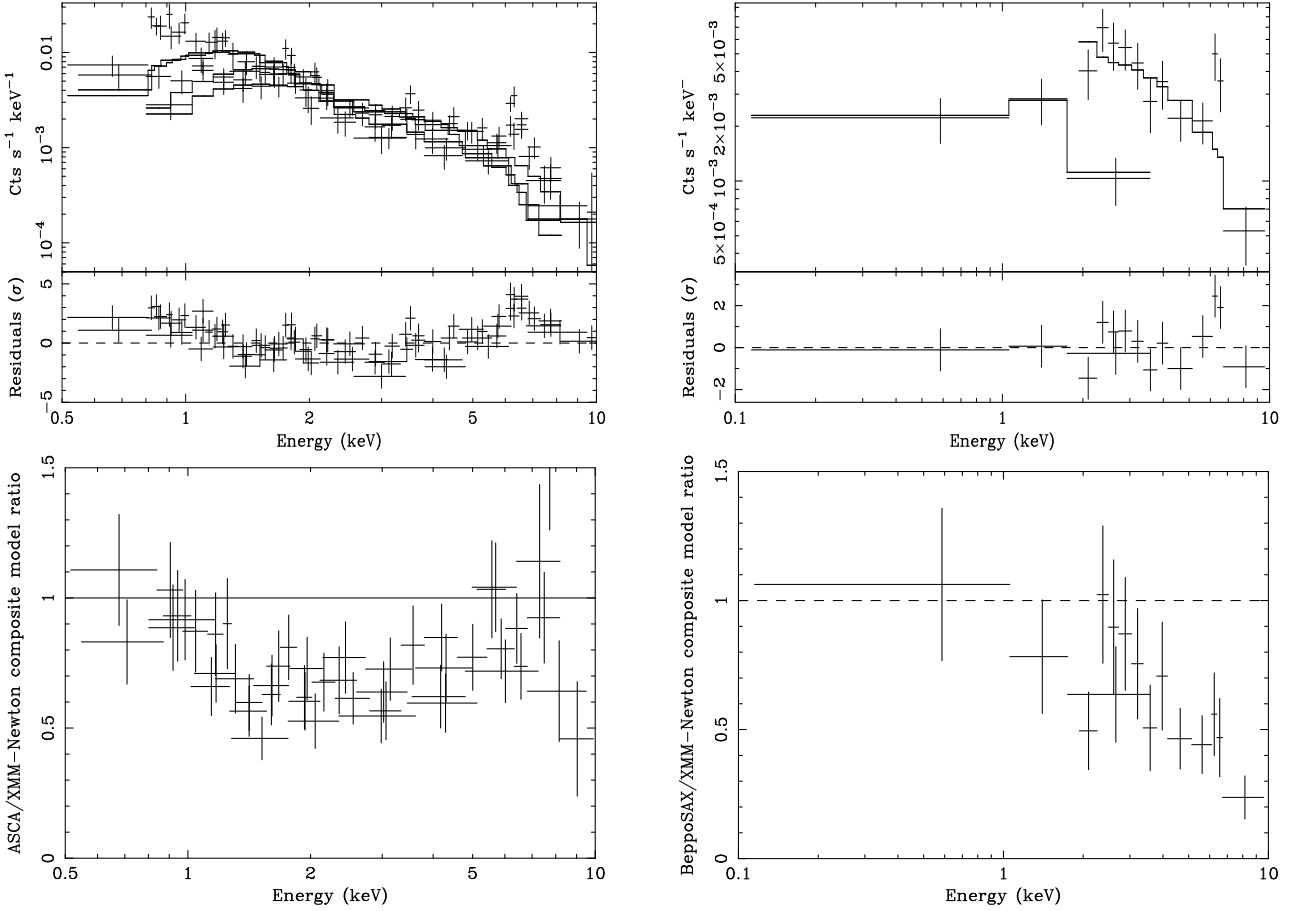

\begin{center}
\hbox{
\psfig{figure=fig6a.ps,height=6.0cm,width=8.0cm,angle=-90}
\hspace{0.75cm}
\psfig{figure=fig6b.ps,height=6.0cm,width=8.0cm,angle=-90}
}
\hbox{
\psfig{figure=fig5a.ps,height=6.0cm,width=8.0cm,angle=-90}
\hspace{0.75cm}
\psfig{figure=fig5b.ps,height=6.0cm,width=8.0cm,angle=-90}
}
\end{center}
\vspace{-0.5cm}
\caption{
{\it Top}:
spectra ({\it upper panels}) and residuals in units of
standard deviation ({\it lower panels}), when
the ASCA
({\it left}) and BeppoSAX ({\it right}) spectra
are fit with a power-law model.
{\it Bottom}: ratio between the ASCA and BeppoSAX
spectra and the XMM-Newton ``composite'' model
(details in text).
}
\label{fig5}
\end{figure*}

The
ASCA and BeppoSAX spectra
exhibit mutually different spectral shapes,
notwithstanding the universal presence of the
large EW K$_{\alpha}$ fluorescent iron line
(see Fig.~\ref{fig5}). A simple power-law
continuum is an adequate representation of
the BeppoSAX spectrum, with a steep
index ($\Gamma = 1.9 \pm^{0.4}_{0.3}$).
This is markedly discrepant with the photon
index measured by XMM-Newton, when a simple
power-law is applied above 2~keV ($\Gamma \simeq
-0.8$; cf. Sect.~2.1.2).
The ASCA spectra requires
at least two continuum components. Adopting
for simplicity the combination of a thermal
and a power-law component, the temperature of
the former is consistent with
$kT_h$ as measured by XMM-Newton, whereas
the latter's spectral index is
intermediate between BeppoSAX and
XMM-Newton ones. A summary of
the ASCA and BeppoSAX best-fits - which represent
for each
case only one of the possible statistically equivalent
solutions - is reported in Table~\ref{tab6}.
\begin{table}
\caption{Best-fit parameters and results for the
ASCA and BeppoSAX spectra of the 3$\arcmin$ radius
region encompassing NGC~5643. The best-fit continua
are: {\tt mekal} plus power-law for ASCA, and
power-law for BeppoSAX, respectively. Both
models include a narrow Gaussian profile, and
absorption by a column density $N_{H,Gal} = 8.3
\times 10^{20}$~cm$^{-2}$.}
\begin{tabular}{lcc} \hline \hline
Parameter & ASCA & BeppoSAX \\
$\Gamma$ & $1.04 \pm 0.14$ & $1.9 \pm^{0.4}_{0.3}$ \\
$kT$ (keV) & $0.68 \pm^{0.10}_{0.09}$ & ... \\
$E_c$ (keV) & $6.45 \pm 0.11$ & $6.43 \pm^{0.10}_{0.16}$ \\
$EW$ (keV) & $1.7 \pm 0.4$ & $1.9 \pm^{1.4}_{1.0}$ \\
0.5--2~keV flux$^a$ & $3.3 \pm^{0.5}_{0.4}$ & $5.4 \pm^{1.6}_{1.4}$ \\
2--10~keV flux$^a$& $9.8 \pm 1.5$ & $10 \pm 2$ \\ 
$\chi^2/\nu$ & 131.0/184 & 9.1/13 \\ \hline \hline
\end{tabular}

$^a$absorption-corrected,
in units of $10^{-13}$~erg~cm$^{-2}$~s$^{-1}$
\label{tab6}
\end{table}

\begin{figure}
\begin{center}
\psfig{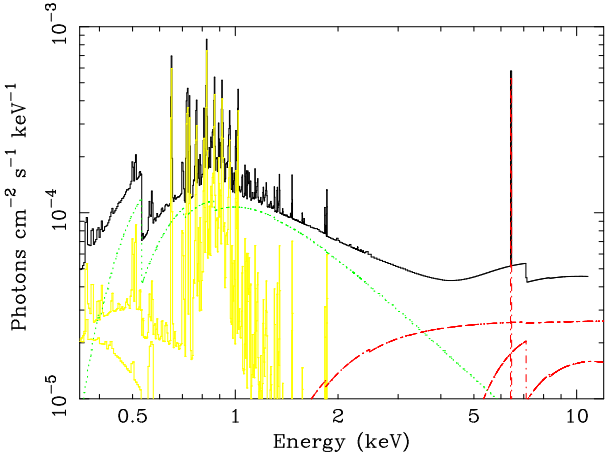}
\end{center}
\vspace{-0.5cm}
\caption{
XMM-Newton ``composite'' spectrum, combining NGC~5643 nucleus
(Model \#3 in Table~\ref{tab2}), and X-1 source best-fits. {\it
Pale solid lines}: thermal components;
{\it dashed lines}: AGN continuum and
K$_{\alpha}$ iron fluorescent line;
{\it dotted line}: X-1 best-fit power-law;
{\it thick, upper solid line}: total composite spectrum
}
\label{fig10}
\end{figure}

In order
to quantitatively estimate the
differences with respect to the XMM-Newton measurements, we have built a
``composite'' XMM-Newton spectrum, combining the best fit models
for the NGC~5643 nucleus and
source X-1. This ``composite'' spectrum is shown in
Fig.~\ref{fig10}, whereas the ratios between the ASCA
and BeppoSAX spectra and the ``composite'' spectrum
are shown in Fig.~\ref{fig5}.
Both these ratios are consistent with 1
below 1~keV. Above 1~keV their behaviors differ:
the ratio against the
BeppoSAX spectrum decrease monotonically up
to 10~keV; the ASCA ratio remains basically constant
around a value of about 0.6, with a local excess
around the iron K$_{\alpha}$ iron line.

In principle, applying this ``composite'' model to the ASCA
and BeppoSAX spectra should provide the ultimate answer to
the question on
which source in the NGC~5643 field is responsible for the flux and spectral
variations observed across the X-ray history of NGC~5643.
Unfortunately,
the composite model has too many parameters,
and would therefore largely overfit the ASCA and BeppoSAX spectra.
We have therefore limited the scope of the exercise,
keeping all the parameters of the composite
model frozen to the XMM-Newton values,
except the continua and iron line normalizations. This
represents of course
a very crude (and admittedly arbitrary) assumption. The
results of this exercise are summarized in Table~\ref{tab8}.
\begin{table}
\caption{Continua and iron line normalizations when
the ``composite'' model is applied to the ASCA and
BeppoSAX spectra. The
XMM-Newton best-fit values are shown as well for
comparison. Legenda:
$N_{pl,a}$: normalization of the locally absorbed
power-law; $N_{pl,u}$: normalization of the
locally unabsorbed
power-law; $N_{th,s}$: normalization of the softer
thermal component; $N_{th,h}$: normalization of the harder
thermal component. The superscripts $^{n}$ and
$^{X-1}$ refers to the NGC~5643 nucleus and
sources X-1, respectively. Errors are at 1-$\sigma$
level for 1 interesting parameter. Values are expressed
as $\times 10^{-5}$ of the corresponding units.
Power-law normalizations are at 1~keV}
\begin{tabular}{lccc} \hline \hline
Normalization & ASCA & BeppoSAX & XMM-Newton \\
$N^n_{pl,a}$ & $7.2 \pm 1.9$ & $< 0.6$ & $8.1 \pm^{0.8}_{1.0}$
\\
$N^n_{pl,u}$ & $0.81 \pm^{0.04}_{0.40}$ & $< 0.5$ & $0.27 \pm 0.02$ \\
$N^{X-1}_{pl,a}$ & $<0.4$ & $1.4 \pm 0.6$ & $1.97 \pm 0.05$ \\
$N^n_{th,s}$ & $4 \pm 3$ & $< 18$ & $2.5 \pm^{0.2}_{0.3}$ \\
$N^n_{th,h}$ & $0.7 \pm 0.2$ & $1.6 \pm^{0.7}_{1.6}$ & $0.83 \pm
0.05$ \\
$I^n_{line}$ & $1.8 \pm 0.3$ & $2.0 \pm^{0.8}_{1.2}$ & $1.6 \pm^{0.3}_{0.2}$ \\ \hline \hline
\end{tabular}
\label{tab8}
\end{table}
Its main conclusions can be outlined as follows:

\begin{itemize}

\item the ASCA soft X-ray faint state might be
mainly due to
source NGC~5463 X-1 becoming at least five times
fainter than during the XMM-Newton (or BeppoSAX)
observation

\item the hard X-ray steep BeppoSAX spectrum might be
mainly due to a decrease of the flux associated with
the nuclear emission, leaving
source X-1 as the dominant contributor to
the hard X-ray flux

\end{itemize}

It is worth noticing that the intensity
of the iron line is consistent with being constant
across all the observations.

Given the crudeness of the above assumption, these
conclusions must be regarded as no more than
hints to the true physical picture. The ultimate answer
on the nature of the X-ray variability in this system
must await future re-observations with
either {\it Chandra} or XMM-Newton.

\section{Discussion}

\subsection{The nature of the obscured AGN in NGC~5643}

A qualitative difference exists between objects obscured
by matter with a column density smaller
(Compton-thin) or larger (Compton-thick) then $N_H 
\simeq \sigma_t^{-1} \simeq 1.5 \times 10^{24}$~cm$^{-2}$
at energies lower than 10~keV, where most of the X-ray
detectors flown so far were sensitive. Compton-thick
AGN can be observed only along optical paths, which
do not intercept the absorbing matter. This may make
highly uncertain the determination of their intrinsic
luminosity, which is dependent on the largely unknown
distribution and covering fraction of the reflecting
and/or scattering matter. Even more importantly,
a Compton-thick absorber substantially
suppresses the incident radiation even at energies
larger than the photoelectric cut-off. At 30~keV
(100~keV), the fraction of transmitted radiation
with respect to an absorber with $N_H = 10^{23.75}$~cm$^{-2}$
is about 85, 30, 4~per cent (50, 20, 2~per cent) for
$N_H = 10^{24.25}$~cm$^{-2}$, $10^{24.75}$~cm$^{-2}$,
$10^{25.25}$~cm$^{-2}$, respectively \cite{wilman99}.
Knowing how many Compton-thick
AGN exist may have an impact on the history
of accretion in the universe \cite{fabian99}

Previous BeppoSAX observations
suggested that Compton-thick
objects represent a large fraction of the Seyfert~2
population \cite{maiolino98}, maybe as large as
30--50~per cent \cite{risaliti99}.
To identify Compton-thick AGN, the measurement
of two critical observables was required:
the shape of the power-law
continuum above 2~keV, and the EW of the K$_{\alpha}$
iron line. Observed flat spectra ($\Gamma \le 1.0$),
and large EW ($\ge$ a few hundreds eV) have been
considered signatures of a reflection-dominated
({\it i.e.}, Compton-thick) AGN.
In principle, a detection by the BeppoSAX PDS above 15~keV
was crucial in determining the nature of
the X-ray absorption (Matt et al. 1997b; Matt et al. 1999).
However, the PDS sensitivity
(a fraction of mCrab in a typical 50~ks observation)
prevented robust detections in most
Seyfert~2 galaxies. Based on the BeppoSAX
results, Maiolino et al. (1998) had classified the
NGC~5643 nucleus as a``warm-scattered" Compton-thick
AGN, obscured by $N_H \ge 10^{25}$~cm$^{-2}$.
This classification was mainly driven by the
steep, unabsorbed spectrum observed by BeppoSAX,
which is inconsistent with
Compton-reflection dominance.
It is interesting to observe that this classification
was not supported by the measurement of the
K$_{\alpha}$ fluorescent iron line centroid, which was
(and has been across the whole X-ray history of
NGC~5643)
inconsistent with He- or H-like iron fluorescent
K$_{\alpha}$ transitions.
XMM-Newton measured
a Compton-thin absorber instead, with
a column density in the range $N_H$6--10$\times
10^{23}$~cm$^{-2}$.
The absorber either directly covers
the nuclear emission, or its Compton-reflection.
In the latter scenario, we might be
observing reprocessing from the
inner far side of the ``torus", partly absorbed
by the outer rim of its near side. This scenario
would require a rather fine-tuned orientation. This,
together with the lack of appropriate statistics,
might explain why this kind of scenario has
never been invoked so far in the context of
reflection-dominated AGN.
Incidentally, the lower column
density helps to substantially reduce (but not to
completely reconcile) the need for a
dust composition geared toward large dust grains
implied by X-ray obscuration
with respect to the observed narrow Br$_{\alpha}$
line component fluxes \cite{lutz02}.

The lack of detection of transmitted nuclear
emission in the Compton-reflection dominated scenario
might be due to the line-of-sight toward the
nucleus intercepting Compton-thick matter.
This may indicate that the ``torus'' is thicker,
the closer to its mid-plane one looks \cite{matt00}.
Alternatively, the line of sight to the nucleus
during the
XMM-Newton observation could be probing a transient
phase of low activity.
The recent serendipitous discovery
of transitions between transmission- to
reflected-dominated spectra of Seyfert~2 galaxies
(see Matt et al. 2003, and references therein) shows
that the classification of an obscured AGN as
Compton-thick/-thin can be time-dependent. In at least
two well studied cases (NGC~2992; Gilli et al. 2000;
NGC~6300, Guainazzi 2002), these transitions are
due to changes of the overall AGN output by more
than one order of magnitude, on timescales of the
order of a few years. The reflection-dominated state
in these cases is the ``echo" of a previous state
of AGN activity, as in the "swan song" state of the
Narrow Line Seyfert~1 galaxy NGC~4051 (Guainazzi et 
al. 1998; Uttley et al. 1999). Interestingly
enough, earlier ASCA and BeppoSAX spectra
of the NGC~5643 nucleus are markedly
different in both flux and spectral shape from those
observed by XMM-Newton. Taking into account the
possible contamination of ``source X-1" into the large
ASCA and BeppoSAX aperture, the observed spectral
variability is consistent with an
historical dynamical range in the AGN power of
one order of magnitude.
Indeed, our interpretation of the steep
spectrum observed by BeppoSAX is that the
nucleus was outshone by NGC~5643~X-1, due to
a phase of very low AGN brightness. This interesting
possibility can be tested with future monitoring
campaigns of this active nucleus, that we plan to
pursue in the nearby future. Of course, we cannot
rule out that the steep BeppoSAX spectrum is due
to a revival of the AGN itself. This would make the
X-ray history of NGC~5643 even more intriguing, though.

The nature of the soft X-ray ({\it i.e.} $\le 2$~keV) emission
in the NGC~5643 nucleus is still uncertain. All
possible models for the soft X-ray spectrum
requires a contribution from emission
lines. The quality of the data is, however, not good enough
to establish the physical process responsible for them. A
good fit is obtained with the composition of two thermal
emission components from optically thin, collisionally ionized
plasma, with temperatures $kT \simeq 0.15$~keV, and
$kT \simeq 0.67$~keV. A possible origin for this emission
is gas heated by shocks or winds in regions of intense
star formation. The 0.5-4.5~keV
luminosity associated with these
thermal components ($L_{0.5-4.5 \ keV} \simeq
1.4 \times 10^{40}$~erg~s$^{-1}$)
is consistent with the empirical
relation discovered by David et al. (1992) with the
Far InfraRed (FIR) luminosity in starburst
galaxies: NGC~5643 $L_{FIR} =
1$--$2 \times 10^{10} L_{\odot} \simeq L_{0.5-4.5 \ keV}^{0.92}$
(Genzel et al. 1998; Moran et al. 2000).
On the other hand, the 2--10~keV luminosity of the
``ultra-hot" thermal component in Model\#2 of
Table~\ref{tab2} ($\simeq 5 \times 10^{39}$~erg~s$^{-1}$)
is only marginally consistent
with the empirical relation between 2--10~keV
and FIR luminosity established by Ranalli et al. (2003).
Conversely, one can estimate the bolometric
luminosity of the
AGN in NGC~5643, $L^{AGN}_{bol} \sim 30 L_X \simeq 7.5 \times
10^{42}$~erg~s$^{-1}$, if it is has a typical Spectral
Energy Distribution as in, {\it e.g.,} Elvis et al. (1994)
and one assumes Model\#3 in Table~\ref{tab2}.
A substantial contribution of AGN reprocessing to the
measured FIR luminosity would not be required. However,
evidence for an intense nuclear
starburst in this galaxy is still controversial.
Alternatively, the EPIC soft X-ray
spectrum can be fit with a pure photoionized
plasma, whose signature in the optical band
would be represented by the bright one-sided
ionization cone \cite{simpson97}. In this scenario,
it is ruled out that a substantial contribution
to the soft X-ray emission comes from scattering
of the nuclear continuum, as this would imply
too steep a slope to be consistent with
that inferred from the hard X-ray spectrum of 
this very same observation. However,
it is possible to fit the soft X-ray spectrum
with a pure emission-line model.
High-resolution spectroscopy observations
of bright Seyfert~2 galaxies have indeed
shown that the contribution of scattered
continuum to the photoionized emission could
be negligible \cite{kinkhabwala02}. In this scenario the
line identification is not fully unambiguous,
and contamination between transitions of different elements likely.
At 0th-order, the interpretations fall into two broad
classes. In the first the spectrum is dominated by
He-like K$_{\alpha}$ transitions of Carbon,
Oxygen, and Neon, with features from the last element being
particularly abundant and including as well
a K$_{\beta}$ line, and a Radiative Recombination
Continuum (RRC). Interestingly enough,
the centroid of the feature encompassing the O{\sc vii} triplet -
which EPIC data cannot resolve - is consistent with
intercombination and resonant lines 
being the dominant contributors of
the
triplet, by contrast to NGC~1068.
This suggests a large density
($n \ge 10^{12}$~cm$^{-3}$) for the photoionized
medium \cite{porquet00}.
In a second possible class of
interpretations, the line spectrum is dominated
by transitions of highly ionized iron species
from Fe{\sc xvii} to Fe{\sc xxiv}. This interpretation
would be consistent with the detection of
a weak Fe{\sc xxvi} K$_{\alpha}$ line in
the hard part of the EPIC spectrum (cf. Fig.~\ref{fig9}).
It is interesting to observe that our
line deconvolution of the EPIC spectra contains all
the brightest lines discovered in the high-resolution
spectrum of NGC~1068 \cite{kinkhabwala02}, except its
O{\sc viii} and N{\sc vii}. O{\sc vii}$\beta$ 
is missing as well, although in this case it might
be blended to invisibility with the N{\sc vii} RRC.

The quest for the physical origin of the soft X-ray
emission in this object requires instrumentation of enough
power to be able to perform high-resolution
spectroscopy in the soft X-ray regime. {\it Chandra}
high-resolution imaging could be important as well in
determining whether the soft X-ray nuclear spectrum is
dominated by diffuse emission, associated to the 15$\arcsec$
ionization cone. AGN photoionized
plasma structures extending on scales
of hundreds parsecs have been observed in
nearby Seyfert~2s (Sako et al. 2000; Bianchi et al. 2003; Iwasawa
et al. 2003). Evidence from the XMM-Newton image
is still inconclusive in this respect.

\subsection{NGC~5643 X-1: an ultrabright ULX?}

\begin{figure*}
\begin{center}
\hbox{
\hspace{-0.20cm}
\psfig{figure=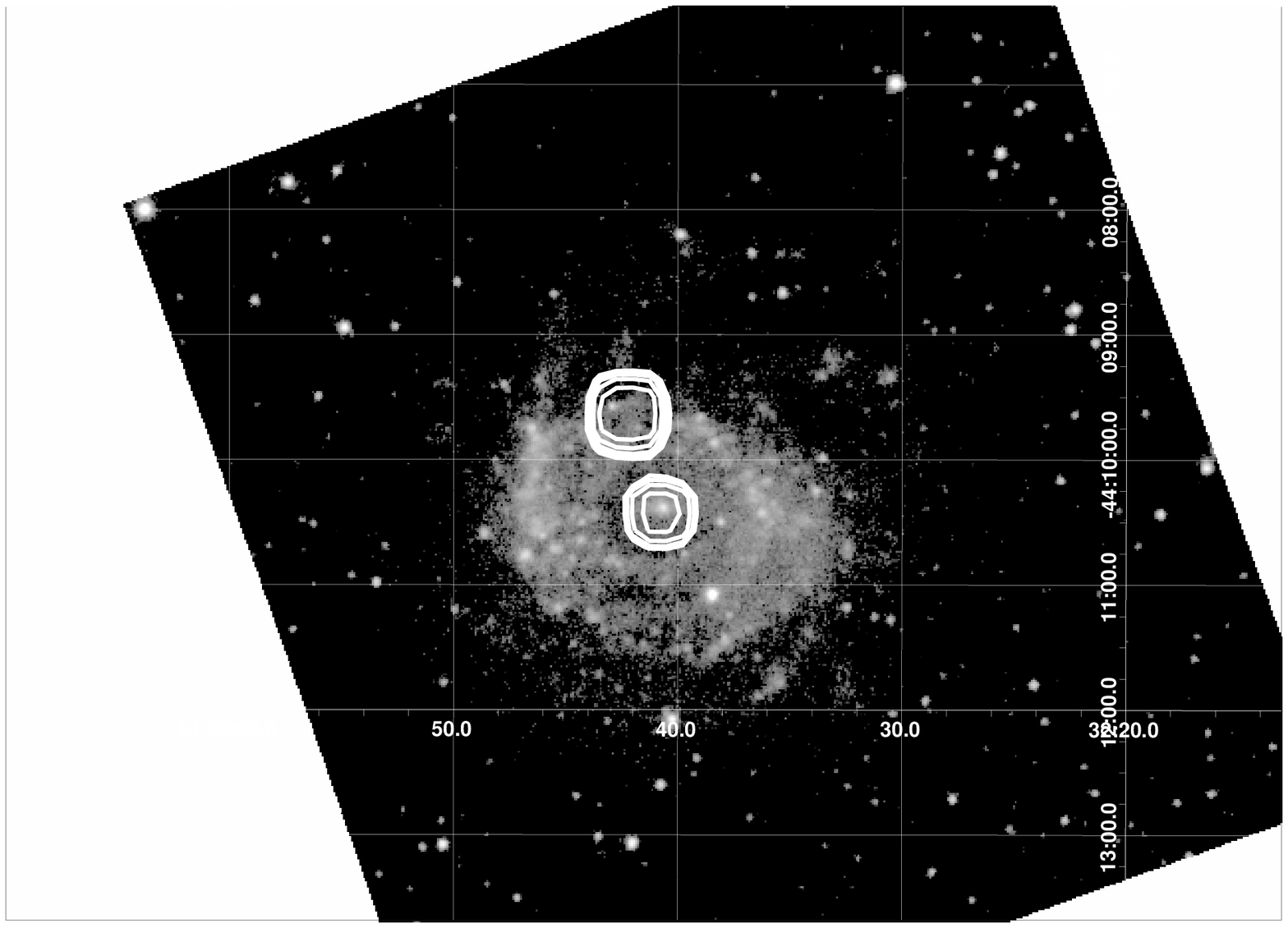,height=8.0cm,width=9.0cm}
\hspace{0.5cm}
\psfig{figure=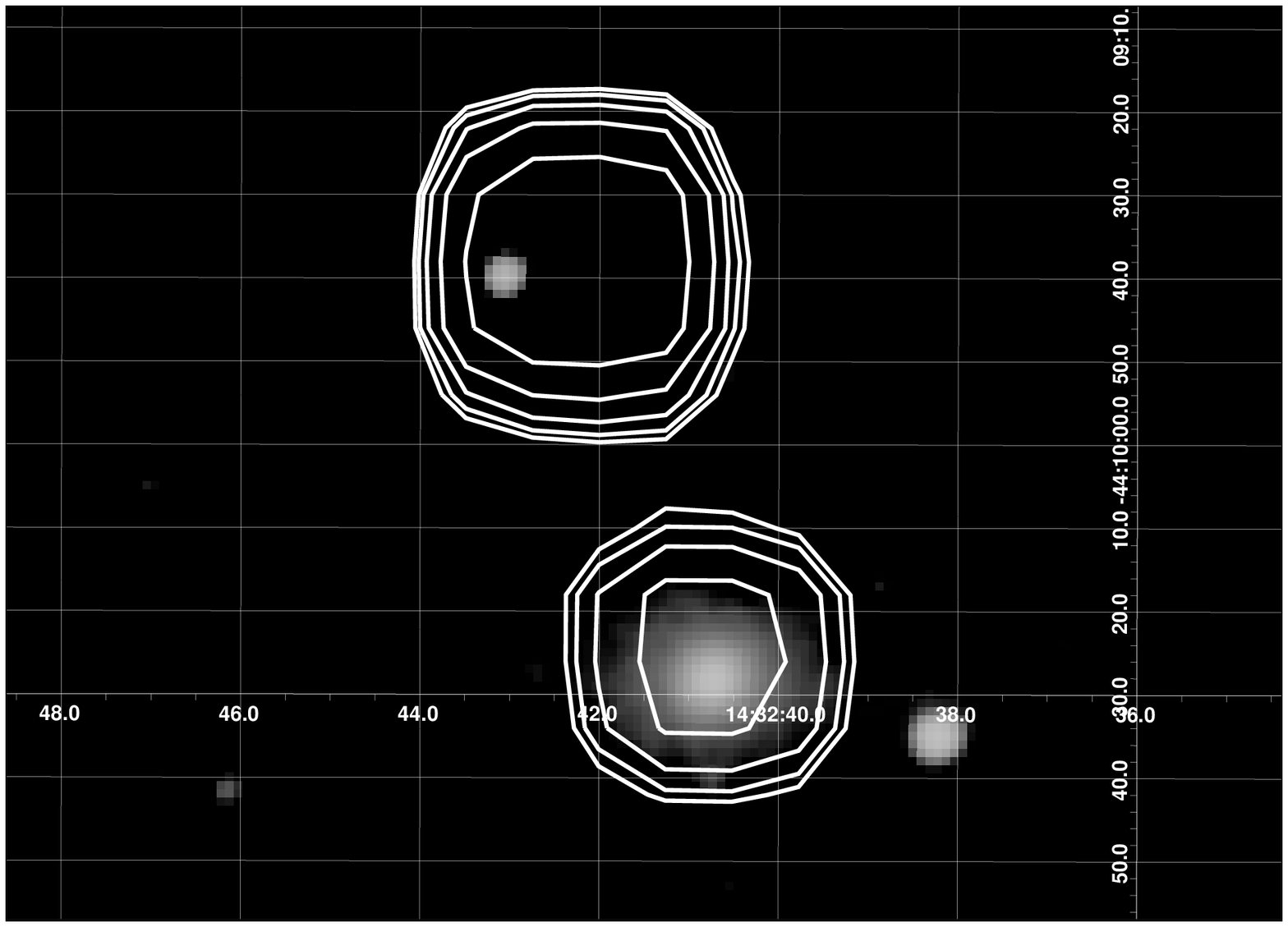,height=8.0cm,width=8.0cm}
}
\end{center}
\vspace{-0.5cm}
\caption{{\it Left panel}
EPIC pn iso-intensity contours
superposed to the innermost science window of the OM/UVW1
exposure. The contours represent 9 logarithmically scaled steps
in the pixel count range between 5 and 169. {\it Right panel}:
zoom of the innermost 1$\arcmin$50$\arcsec$. The pn
iso-intensity contours are superposed to the K-band 2MASS
image
}
\label{fig2}
\end{figure*}

The XMM-Newton image of the NGC~5643 field clearly shows a
source, about 0.8$\arcmin$ NE from the position
of the optical nucleus.
This source - labeled
``NGC~5643 X-1" in this paper - is about a factor 50~per cent brighter
than the nucleus itself, and had been previously observed
by the ROSAT HRI, in a three times fainter state.
In Fig.~\ref{fig2} we show an overlap between the
EPIC intensity contours and the simultaneous
OM image. Although
the relatively broad Point
Spread Function of the XMM-Newton optics prevents
us from pinpointing an unambiguous optical counterpart for
NGC~5643 X-1, it is apparently located at the
outskirts of the host galaxy optical
extension. If the source is not a background
object, its remarkable X-ray luminosity ($\simeq
4.4 \times 10^{40}$~erg~s$^{-1}$) would convert it
in the third brightest ULX ever observed in hard X-rays (Foschini et al.
2002; Humphrey et al. 2003; Swartz et al. 2003).
Its spectral properties
provide relatively little information on its nature.
A fit with a simple power-law ($\Gamma \simeq
1.7$) or bremsstrahlung ($kT \simeq 8$~keV)
continuum with moderate absorption ($N_H \simeq
1$--2$\times 10^{21}$~cm$^{-2}$) is a good
description of the observed spectrum. It is
interesting to observe that a fit
with a multitemperature disk blackbody is poor.

In the quest for the nature of NGC~5643~X-1 the discovery of a
possible counterpart at other wavelengths
would be crucial. No known radio
source can be associated with the position
of NGC~5643~X-1. A 2MASS bright optical source
with coordinates: $\alpha_{2000}$~=~$14^h32^m43.0^s$;
$\delta_{2000}$~=~$-44^{\circ}09{\arcmin}40{\arcsec}$,
about a factor of 3 fainter than the optical nucleus,
is only marginally consistent with typical
uncertainties in XMM-Newton attitude reconstruction.
The brightest OM UV source in the
EPIC error box has coordinates
$\alpha_{2000}$~=~$14^h32^m42.9^s$;
$\delta_{2000}$~=~$-44^{\circ}09{\arcmin}34{\arcsec}$.
The identification of an unambiguous
counterpart needs to await a more precise determination
of the X-ray centroid, which could be achieved with a
{\it Chandra} observation of the galaxy field.
On the other hand, the study of the X-ray history
of the NGC~5643 field suggest that NGC~5643 X-1
underwent flux variations with a dynamical range of
at least 5, thus ruling out an association with
a supernova remnant. Again, future monitoring campaigns
of this field may offer important clues on its
variability pattern, and henceforth on its nature.

\section*{Acknowledgments}

This work is based on observations obtained with XMM-Newton, and ESA
science mission with instruments and contributions directly funded by ESA
Member States and the USA (NASA). This research has made use of the
NASA's Astrophysics Data System Abstract Service, and
of the NASA/IPAC Extragalactic Database (NED),
which is operated by the Jet Propulsion Laboratory,
California Institute of Technology, under contract with the
National Aeronautic and Space Administration. Last, but
not least, we acknowledge careful reading and
useful suggestions from an
anonymous referee.

\end{document}